\documentstyle[aps,prl,preprint,floats,epsf]{revtex} %For preprints, drafts

\begin{document}
\draft
%\tighten
%\twocolumn
\title{{\hfill{CLNS 97/1533}}\\
Improving Measurements of the CKM 
Phase $\gamma$ Using Charm Factory Results}
\author{Abner Soffer}
% e-mail: abi@mail.lns.cornell.edu
 
\address{Laboratory of Nuclear Science, Cornell University, Ithaca, NY 14850, 
USA}

\date{\today}
\maketitle

\begin{abstract}
Several authors have proposed methods to measure the phase $\gamma$ of
the Cabibbo-Kobayashi-Maskawa (CKM) unitarity triangle using decays
of the type $B\rightarrow DK$. We show how to remove uncertainties
from these measurements and increase their sensitivity by
measuring CP-conserving phases at a charm factory.
\end{abstract}

\pacs{PACS numbers:13.25.Hw,14.40.Nd}

\begin{figure}[p]
\centering
\epsfxsize=3.25in
\epsffile{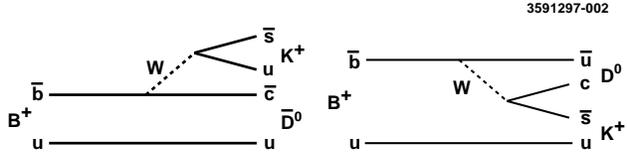}
\caption{Dominant Feynman diagrams of $B^+\rightarrow \overline D^0 K^+$ and 
	$B^+\rightarrow D^0 K^+$.}
\label{fig:dk-diagrams}
\end{figure}

\begin{figure}[p]
\centering
\epsfxsize=3.25in
\epsffile{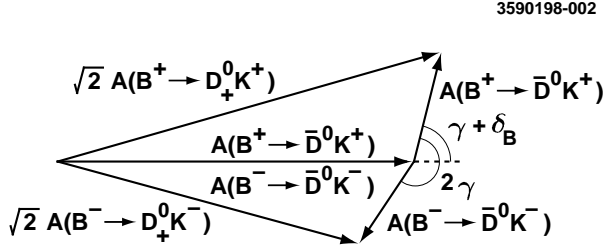}
\caption{The triangle relation
$D^0_+ \equiv {1\over\sqrt{2}} \left(D^0 + \overline D^0\right)$
applied to the $B^+$ and $B^-$~decay amplitudes. Similar triangles
exist for $D^0_- \equiv {1\over\sqrt{2}} \left(D^0 - \overline D^0\right)$.}
\label{fig:GLW-triangle}
\end{figure}

CP-violation experiments can be used to test the standard model and
probe new physics. Decays of the type $B\rightarrow DK$ provide
several possibilities for carrying out such tests.  Gronau, London and
Wyler (GLW)~\cite{ref:GLW} have shown that the CKM~\cite{ref:ckm}
phase $\gamma \approx \arg(V_{ub}^*)$ can be measured in the
interference between the decays $B^+\rightarrow \overline D^0 K^+$ and
$B^+\rightarrow D^0 K^+$, whose Feynman diagrams are shown in
Figure~\ref{fig:dk-diagrams}. The phase difference between the two
amplitudes is $\gamma + \delta_B$, where $\delta_B$ is a
CP-conserving, final state interaction (FSI) phase. Interference
occurs when the $D$ is observed as one of the CP-eigenstates
\begin{equation}
D^0_\pm \equiv {1\over\sqrt{2}} \left(D^0 \pm \overline D^0\right),
	\label{eq:GLW-triangle}
\end{equation}
which are identified by their decay products. Disregarding CP-violation in
$D$~decays and $D^0 - \overline D^0$ mixing,
Equation~(\ref{eq:GLW-triangle}) implies the triangle relations of
Figure~\ref{fig:GLW-triangle}, from which $\gamma$ is extracted. As
with several other methods, $\gamma$ is obtained up to a four-fold
discrete ambiguity. This method can be applied to any decay mode of
the type $B^+\rightarrow D^0_\pm K^+ n\pi$, and in principle does not
require $\delta_B$ to be different from zero.

Several variations of the GLW~method have been
developed. Dunietz~\cite{ref:dunietz} has suggested applying the method to
$B^0 \rightarrow D^0_\pm K^{*0}$, where the subsequent decay $K^{*0}
\rightarrow K^+\pi^-$ tags the flavor of the $B$. In this mode both
amplitudes are color suppressed, and thus have similar
magnitudes. This results in greater interference than one expects in
$B^+\rightarrow D^0_\pm K^+$, where the the $\overline b\rightarrow
\overline c$ amplitude has a color allowed contribution, while the
$\overline b\rightarrow \overline u$ amplitude is color
suppressed. Atwood, Eilam, Gronau and Soni(AEGS)~\cite{ref:resonance}
have considered the decays $B^+\rightarrow D^0_\pm [K^+]$, where
$[K^+]$ is a state such as $K\pi$ or $K^*\pi$, with invariant mass
around 1400~MeV, where several strange resonances can interfere. The
different Breit-Wigner widths of the resonances lead to large and
calculable CP-conserving phases, enabling the extraction of $\gamma$
from decay rate asymmetries, in addition to using the GLW triangle
method.

Atwood, Dunietz and Soni(ADS)~\cite{ref:ads} have pointed out a serious
difficulty with the GLW~method, originating from the fact that in
order to measure the branching fraction ${\cal B}(B^+\rightarrow
D^0K^+)$, the $D^0$ must be identified in a hadronic\footnote{Full
reconstruction is impossible in semileptonic decays, resulting in
unacceptably high background levels.} final state, $K^-\pi^+ n\pi$. To
simplify the discussion, we explicitly refer to $K^-\pi^+$. ADS noted
that the decay chain $B^+\rightarrow D^0 K^+$, $D^0\rightarrow
K^-\pi^+$ results in the same final state as $B^+\rightarrow \overline
D^0 K^+$, $\overline D^0\rightarrow K^-\pi^+$, where the $\overline
D^0$ undergoes doubly Cabibbo suppressed decay. Using only measured
quantities and factorization, they estimated the ratio between the
interfering decay chains:
\begin{eqnarray}
&&\left|{A(B^+\rightarrow \overline D^0 K^+) \; 
	A(\overline D^0\rightarrow K^- \pi^+)
	\over
A(B^+\rightarrow D^0 K^+) \; A(D^0\rightarrow K^- \pi^+)}\right|\nonumber\\
 &\approx&
	\left|{V_{cb}^* \over V_{ub}^*}  \;
	{V_{us}  \over V_{cs}}\;
	{a_1 \over a_2}\right| \;
	\sqrt{{\cal B}(\overline D^0 \rightarrow K^-\pi^+) \over 
		{\cal B}(D^0 \rightarrow K^-\pi^+)}
	\nonumber\\[5pt]
	&\approx& {1 \over 0.08}\; 0.22 \;
	{1\over 0.26} \;
	\sqrt{0.0077} 
	\approx 0.9. \label{eq:GLW-trouble}
\end{eqnarray}
Evidently, sizable interference makes it practically impossible to
measure ${\cal B}(B^+\rightarrow D^0K^+)$, and the GLW~method
fails. For $B^0 \rightarrow D^0_\pm K^{*0}$,
Equation~(\ref{eq:GLW-trouble}) yields the ratio 0.25. The problem is
thus less severe in this mode, but it nevertheless introduces a
significant systematic uncertainty to the measurement of $\gamma$.

ADS proposed to use this interference to obtain $\gamma$ from the
decay rate asymmetries in $B^+\rightarrow [f_i]K^+$, where $f_i, \ i =
1,2,$ are two different hadronic states, at least one of which is of
the type $K^-\pi^+ n\pi$, which can arise from the Cabibbo allowed
decay of a $D^0$ or the doubly Cabibbo suppressed decay of a
$\overline D^0$. Measuring the four branching fractions
\begin{equation}
{\cal B}(B^+ \rightarrow [f_i] K^+), \ \ \ \ \ \
{\cal B}(B^- \rightarrow [\overline f_i] K^-),
\label{eq:ads-brs}
\end{equation}
one calculates the four unknowns which cannot be measured in other
$B$~decays, namely ${\cal B}(B^+\rightarrow D^0 K^+)$, $\gamma$ (up to
a discrete ambiguity), and the CP-conserving phase differences of the
two final states. For a given final state $[f]$, the CP-conserving
phase is
\begin{equation}
\delta = \delta_B + \delta_D,\label{eq:deltas}
\end{equation}
where 
\begin{equation}
\delta_D = {\rm arg}\left[A(D^0\rightarrow [f])
			    A(\overline D^0\rightarrow [f])^*\right]
	\label{eq:delta_d}
\end{equation}
can be quite large when $[f]$ is a flavor
eigenstate~\cite{ref:ddecays}, resulting in CP-asymmetries of order
100\%.

ADS noted that since $\delta_B$ is identical for all $D$~decay modes,
an independent, charm factory measurement of $\delta_D$ would add a
constraint to Equation~(\ref{eq:deltas}), increasing the sensitivity
of the method.  The need to use two decay modes still causes
significant loss of statistical power, however, since the only modes
for which the product of the efficiency and branching fraction is
sizable are $K^-\pi^+$, $K^-\pi^+\pi^0$ and $K^-\pi^+\pi^-\pi^+$.  The
benefits of the $\delta_D$ measurements are much greater, however, if
one makes the additional hypothesis
\begin{equation}
\delta_B \approx 0. \label{eq:fsi}
\end{equation}
It has recently been shown~\cite{ref:donoghu} that rescattering phases
should not be strongly suppressed in $B$~decays, contrary to previous
expectations~\cite{ref:bjorken}. Tight experimental limits on FSI
phases in $B\rightarrow D\pi$, $D^*\pi$, $D\rho$ and
$D^*\rho$~\cite{ref:fsi}, however, provide good indication that the
phases could be small in two~body $B\rightarrow D^{(*)} + {\rm light \
hadron}$ decays. Tests of this hypothesis specifically in the case of
$B\rightarrow DK$ are discussed below.

\begin{figure}[p]
\centering
\epsfxsize=3.25in
\epsffile{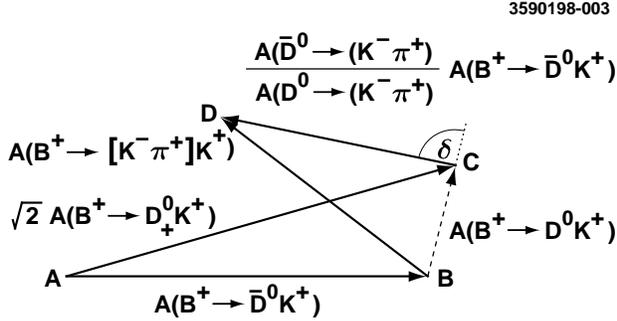}
\caption{Solution of the problem with the GLW~method: Given
$\delta$, {\it BD}\/ and {\it CD},\/ the length {\it
BC}\/ is determined, enabling the construction of the triangle {\it ABC}.}
\label{fig:GLW-triangle-improved} 
\end{figure}

Given Equation~(\ref{eq:fsi}), an independent measurement of the phase
$\delta_D$ would enable the extraction of $\gamma$ and ${\cal
B}(B^+\rightarrow D^0 K^+)$ from the branching fractions of
Equation~(\ref{eq:ads-brs}) using a single flavor eigenstate $D$~decay
mode, provided $\sin\delta_D$ is not too close to 0.  This would free
other modes to be used for increasing statistics and resolving
discrete ambiguities. In addition, we note that knowledge of $\delta$
solves the problem which Equation~(\ref{eq:GLW-trouble}) implies for
the GLW~method. The method requires measuring the lengths of all sides
of the triangle {\it ABC}\/ of Figure~\ref{fig:GLW-triangle-improved}.
As discussed above, the side {\it BC}\/ cannot be measured directly,
since ${\cal B}(B^+ \rightarrow [K^- \pi^+]K^+)$ actually measure {\it
BD},\/ which is the interference of {\it BC}\/ with {\it CD}, \/ the
$\overline b\rightarrow \overline c$ transition amplitude followed by
a doubly Cabibbo suppressed $D$~decay. However, given $\delta$,{\it
CD}\/ and {\it BD},\/ the length {\it BC}\/ is determined. The
triangle {\it ABC}\/ is now fully constructed, and $\gamma$ is
obtained using the GLW~method.

\begin{figure}
\centering
\epsfxsize=3.25in
\epsffile{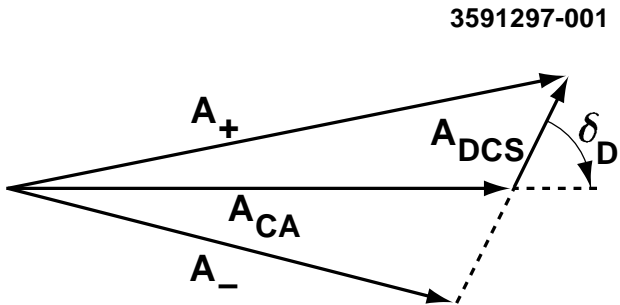}
\caption{Obtaining the phase $\delta_D$ of Equation~(\ref{eq:delta_d}) from the
	Cabibbo allowed $D^0$~decay amplitude, $A_{CA}$, the doubly Cabibbo
	suppressed amplitude, $A_{DCS}$, and their interference,
	$A_\pm \equiv A_{CA}\pm A_{DCS}$.}
\label{fig:3770-triangle}
\end{figure}

We proceed to study the measurement of $\delta_D$ at a charm factory,
operating at the $\psi(3770)$ resonance.  Equation~(\ref{eq:delta_d})
is graphically represented in Figure~\ref{fig:3770-triangle},
demonstrating how to obtain $\delta_D$ from the the Cabibbo allowed
$D$~decay amplitude, $A_{CA}$, the doubly Cabibbo suppressed
amplitude, $A_{DCS}$, and their interference, $A_\pm \equiv A_{CA} \pm
A_{DCS}$.  While $A_{CA}$ and $A_{DCS}$ have been measured at
CLEO~\cite{ref:double-cabibbo} for the $K^-\pi^+$ mode by using
$D^{*+}$ decays to tag the $D^0$~flavor, measuring $A_\pm$ requires
producing $D^0\overline D^0$ pairs in a known coherent state.  It is
therefore best to perform all three measurements at the charm factory,
canceling many systematic errors in the construction of the triangles
of Figure~\ref{fig:3770-triangle}. To measure the amplitude $A_+$
($A_-$), one of the $\psi(3770)$ daughters is tagged as a $D^0_-$
($D^0_+$) by observing it decay into a CP-odd (CP-even) state, such as
$K_S\;\pi^0$ ($K^+K^-$). The other daughter is then $D^0_+$ ($D^0_-$),
and the fraction of the time that it is seen decaying into $K^-\pi^+$
gives the interference amplitude ${1\over2}|A_\pm|^2$. We immediately
find
\begin{equation}
\cos\delta_D = \pm {|A_\pm|^2 - |A_{CA}|^2 - |A_{DCS}|^2 \over
	2 |A_{CA}| | A_{DCS}|}.
\end{equation} 
Due to the low statistics tagging scheme of the $|A_\pm|$ measurement
and the fact that $\left|A_{DCS}\right| \ll \left|A_{CA}\right|$, the
error in $\cos\delta_D$ is dominated by the $|A_\pm|$ measurement
error. Hence
\begin{eqnarray}
\Delta \cos\delta_D
	&\approx& {\Delta |A_\pm| \over \left|A_{DCS}\right|}
	\approx {\Delta |A_\pm| \over |A_\pm|} \;
		  \left|{A_{CA} \over A_{DCS}}\right| \nonumber\\
	&=&
		{1 \over 2 \sqrt{N_{A_\pm}}} \;
		  \left|A_{CA} \over A_{DCS}\right|, \label{eq:cos-delta-err2}
\end{eqnarray}
where $N_{A_\pm}$ is the number of events observed in the $A_\pm$
channels, and we made use of $\left|A_{CA}\right| \approx
\left|A_\pm\right|$.  Since the event is fully reconstructed,
background is expected to be small, and its contribution to $\Delta
\cos\delta_D$ is neglected in this discussion. $N_{A_\pm}$ is given by
\begin{eqnarray}
	N_{A_\pm} &=&
	N_{D \overline D}
	\ {\cal B}(D^0\rightarrow K^-\pi^+) \; \epsilon(K^-\pi^+)\nonumber\\
	&\times& \sum_i {\cal B}(D^0 \rightarrow t_i) \; \epsilon(t_i),
	\label{eq:num-ddbar-events}
\end{eqnarray}
where $N_{D\overline D}$ is the number of $\psi(3770)\rightarrow D^0
\overline D^0$ events, $t_i$ are the CP-eigenstates used for tagging,
and $\epsilon(X)$ is the reconstruction efficiency of the state
$X$.  From the list of $D^0 \rightarrow$CP-eigenstate branching
fractions and efficiencies in Table~\ref{tab:tagging-modes} we obtain
$\sum_i {\cal B}(D^0 \rightarrow t_i) \; \epsilon(t_i) \approx 0.02$.
Taking ${\cal B}(D^0\rightarrow K^-\pi^+) \approx 0.04$,
$\epsilon(K^-\pi^+) \approx 0.8$ and $\left|A_{CA} / A_{DCS}\right| =
1/\sqrt{0.0077}$, Equation~(\ref{eq:cos-delta-err2}) becomes
\begin{equation}
\Delta\cos\delta_D \approx {230 \over \sqrt{N_{D\overline D}}}.
\label{eq:num_psi3770}
\end{equation}
It is expected that in one year the charm factory will collect 10~$\rm
fb^{-1}$, or $N_{D\overline D} =
2.9\times10^{7}$~\cite{ref:charm-fact}, resulting in $\Delta
\cos\delta_D \approx 0.04$.  $\cos\delta_D$ can thus be measured with
more than sufficient precision, even in the presence of background and
with relatively modest luminosity.  We note that the same measurement
technique can be used with multi-body $D^0$~decays, in which
$\cos\delta_D$ varies over the available phase space.  While the
relative statistical error in every small region of phase space will
be large, its effect on the measurement of $\gamma$ in $B\rightarrow
DK$ will be proportionally small. The total error in $\gamma$ due to
$\Delta \cos\delta_D$ will be as small as in the two~body $K^-\pi^+$
mode, up to differences in $D^0$ branching fractions, reconstruction
efficiencies and backgrounds.

Several tests of the hypothesis~(\ref{eq:fsi}) can be conducted.
First, the decay rate asymmetry in $B^+\rightarrow D_\pm^0 K^+$ is
proportional to $\sin\delta_B\sin\gamma$, hence its measurement can be
used to put a limit on $\sin\delta_B$. The best limit should come from
$B^0\rightarrow D_\pm^0 K^{*0}$, where factorization predicts the
ratio between the interfering amplitudes to be $\sim 2.8$, compared
to $\sim 11$ in $B^+\rightarrow D_\pm^0 K^+$.  In addition, since
both $B^0\rightarrow D^0 K^{*0}$ and $B^0\rightarrow \overline D^0
K^{*0}$ proceed through color suppressed amplitudes, one expects the
relation
\begin{equation}
{{\cal B}(B^0\rightarrow D^0 K^{*0}) \over 
	{\cal B}(B^0\rightarrow \overline D^0 K^{*0})}
	\approx \left|{V_{ub}^* \over V_{cb}^*} \; 
	  {V_{cs} \over V_{us}}\right|^2
\end{equation}
to hold well. Agreement between this expectation and the length {\it
BC}, as obtained from the triangle {\it BCD} of
Figure~\ref{fig:GLW-triangle-improved} (replacing $B^+ \rightarrow
B^0$, $K^+\rightarrow K^{*0}$ and applying Equation~(\ref{eq:fsi})),
would support our theoretical picture of such decays, including
the hypothesis~(\ref{eq:fsi}).  Disagreement, on the other hand, would be
strong evidence against this hypothesis.

AEGS have also shown how to
test Equation~(\ref{eq:fsi}) in $B^+\rightarrow D^0_\pm [K^+]$ decays
using the sum of the decay widths $\Sigma = d^2[\Gamma(B^+\rightarrow
D^0_\pm [K^+]) + \Gamma(B^-\rightarrow D^0_\pm [K^-])]/dsdz$, where
$z$ is the $[K^+]$ decay angle and $s$ is the difference between the
invariant mass of the $[K^+]$~state and a weighted average mass of the
interfering resonances. If $\delta_B \approx 0$, then $\Sigma$ is an
even function of $s$. Otherwise, an $s$-odd term is introduced.

Finally, assuming Equation~(\ref{eq:fsi}) , the values of $\delta_D$
measured at the charm factory can be applied to the measurement of
$\gamma$ in all the $B\rightarrow DK$ modes and methods discussed
above. Inconsistencies among the different $\gamma$ measurements would
thus invalidate Equation~(\ref{eq:fsi}).

To summarize, we have shown that the problem with the Gronau-Wyler
method is solved by conducting relatively low-luminosity phase
measurements at a charm factory. This will enable the exploitation of more
$B$~decay modes for the measurement of $\gamma$ than  otherwise
possible, and enhance the sensitivity of the
Atwood-Dunietz-Soni method. We have also indicated how to test the
hypothesis, which our analysis depends on, that the FSI phase is
negligible in $B\rightarrow DK$.

I would like to thank K.~Berkelman, D.~Cassel, M.~Gronau, G.P.~Lepage,
D.~London and H.N.~Nelson for fruitful discussions and suggestions.
This work is supported by the National Science Foundation.

\begin{table}
\begin{center}
\begin{tabular}{lccc}
$t$ & ${\cal B}(D^0\rightarrow t)$ & $\epsilon(t)$ & 
	${\cal B} \times \epsilon$ \\
\hline
$K_S \; \pi^0$  & 0.0106 & 0.3 & 0.003  \\
$K_S \; \eta(\rightarrow \gamma\gamma) $  & 0.007  & 0.1 & 0.0007 \\
$K_S \; \rho^0 $  & 0.012  & 0.4 & 0.005  \\
$K_S \; \omega(\rightarrow \pi^+\pi^-\pi^0)$ & 0.021  & 0.2 & 0.004  \\
$K_S \; \eta' (\rightarrow \pi^+\pi^-\eta) $ & 0.017  & 0.1& 0.0017 \\
$K_S \; \eta' (\rightarrow \rho^0\gamma) $ & 0.017  & 0.1& 0.0017 \\
$K^+  K^- $  & 0.004  & 0.8 & 0.0032 \\
$\pi^+  \pi^- $  & 0.0015  & 0.8 & 0.0012 \\
\hline
total             &          &     &    0.021 \\
\end{tabular}
\end{center}
\caption{Branching fractions~\protect\cite{ref:pdg} of $D^0$ decays to
CP-eigenstates, reconstruction efficiencies, and their
products. Efficiencies include sub-mode branching fractions, such as
$K_S \rightarrow \pi^+\pi^-$, and are constructed assuming 90\% track
and photon efficiency and 50\% $\pi^0$ efficiency.}
\label{tab:tagging-modes}
\end{table}

\begin{references}
\bibitem{ref:GLW} M.~Gronau and D.~London, Phys. Lett. {\bf B253}, 483 (1991); 
	M.~Gronau and D.~Wyler, Phys. Lett. {\bf B265}, 172 (1991).
\bibitem{ref:ckm} M.~Kobayashi and K.~Maskawa, Prog. Theor. Phys. {\bf 49},
	652 (1973).
\bibitem{ref:dunietz} I.~Dunietz, Phys. Lett. {\bf B270}, 75 (1991).
\bibitem{ref:resonance} D.~Atwood, G.~Eilam, M.~Gronau and A.~Soni, 
	Phys. Lett. {\bf B341}, 372 (1995). 
\bibitem{ref:ads} D.~Atwood, I.~Dunietz and A.~Soni, Phys. Rev. Lett. {\bf 78},
	3257 (1997).
\bibitem{ref:ddecays} See, for example, P.I.~Frabetti (E687 Collaboration),
	Phys. Lett. {\bf B331}, 217 (1994).
\bibitem{ref:donoghu} J.F.~Donoghue, E.~Golowich, A.A.~Petrov and
	 J.M.~Soares, Phys. Rev. Lett. {\bf 77}, 2178 (1996).
\bibitem{ref:bjorken} J.D.~Bjorken, Nucl. Phys. B (Proc. Suppl.) {\bf 11},
	325 (1989).
\bibitem{ref:fsi} H.~Yamamoto , CBX~94-14, HUTP-94/A006; H.N.~Nelson,
	private communication.
\bibitem{ref:double-cabibbo} D.~Cinabro {\it et al} (CLEO collaboration), 
	 Phys. Rev. Lett. {\bf 72}, 1406 (1994).
\bibitem{ref:charm-fact} J.~Kirkby, in La Thuile 1996, Results and 
	Perspectives in Particle Physics, 747 (1996). 
\bibitem{ref:pdg} R.M.~Barnett {\it et al.}, (Particle Data Group),
	Phys. Rev. {\bf D54}, 1 (1996).
\end{references}
\end{document}